# WEB-BASED ENTERPRISE INFORMATION SYSTEMS DEVELOPMENT: THE INTEGRATED METHODOLOGY

**Name:** Dr. Sergey V. Zykov, Ph.D.

**Organization:** ITERA Oil and Gas Company L.L.C., Moscow, Russia

**Position:** Internet Project Manager

**Email:** [szykov@itera.ru](mailto:szykov@itera.ru)

**Phone:** +7 (095) 411-8500 ext. 44-21

**Fax:** +7 (095) 411-8502

**Internet:** [http://www.iteragroup.com](http://www.iteragroup.com)

Abstract: The paper considers software development issues for large-scale enterprise information systems (IS) with databases (DB) in global heterogeneous distributed computational environment. Due to high IT development rates, the present-day society has accumulated and rapidly increases an extremely huge data burden. Manipulating with such huge data arrays becomes an essential problem, particularly due to their global distribution, heterogeneous and weak-structured character. The conceptual approach to integrated Internet-based IS design, development and implementation is presented, including formal models, software development methodology and original software development tools for visual problem-oriented development and content management. IS implementation results proved shortening terms and reducing costs of implementation compared to commercial software available.

# 1. INTRODUCTION

State-of-the-art IS and DBMS are currently operating terabyte information arrays, which is going to change for petabyte size DB shortly. Giant heterogeneous data volumes demand totally new methodologies of software design and implementation as well as new tools for continuous IS iterative development and DB and meta-DB (MDB) integration.

Under rapidly growing IT penetration into each and every activity of the modern society, software integration becomes a critical issue, particularly considering co-existing heterogeneous (and often contradicting) concepts, methodologies, models and approaches to handling data. Quite a number of research groups and major software vendors focus on uniform theoretical languages and SDK solutions for enterprise-scale IS design, implementation and integration. However, the problem is still far from its adequate solution.

The above considerations require a conceptually new approach to integrated IS design, implementation and support, that provides dynamic tracking (meta)data integrity and high IS fault tolerance in a heterogeneous computational environment. Therewith, the aim of the research is building a uniform methodology for large-scale IS design, implementation and support for global, distributed, heterogeneous computational environment, which comprises formal methods and a software development toolkit.

# 2. THE METHODOLOGY
## 2.1. Software development model

The suggested conceptual approach to IS design, implementation and support is based on (meta)data integration. One of its essential components is software conceptual design, which provides a unified (meta)data computational model in the form of (M)DO, language and SDK. Another essential ingredient of the approach is the methodology that supports continuous multi-level iterative software design and implementation (with reengineering) from problem domain entities to (meta)data and IS schemes. Therewith, (meta)data actuality, completeness, consistency and integrity are fully controlled throughout the entire IS lifecycle.

Application development requires conceptual and methodological generalization of (M)DO integrated management on the basis of unified, open and extendable interface, language and software tools.

Research methods for the methodology are based on a creative synthesis of finite sequence theory [1], category theory [2,3], computations theory [5] and semantic networks theory [4]. However, all of the approaches known as yet either have methodological "gaps" or do not result in enterprise-level solutions with practically applicable implementation features.

The developed data models (DM) for problem domains and software development tools manipulate dynamic and static features of heterogeneous weak-structured environments more comprehensively. A novel architectural solution is given for an open, extendable, distributed, interoperable, heterogeneous environment that supports personalized front-end and back-end (meta)data processing based on component technology with embedded script procedures.

The variable domain-based DM with states features event-driven DO and MDO control of heterogeneous problem domains. Therewith, the supported range of possible (meta)data sources is extended up to virtually arbitrary data warehouses (including (M)DB), which integrate state-of-the-art front-end architectures of Web-IS, intermediate and legacy systems. The software development solution features content-oriented (M)DM, which is modeled by an abstract machine. As far as implementation part is concerned, traditional technologies and software development tools are extended by continuous conceptual iterative integrated IS design based on a creative combination of COM model, UML language and BPR technology.

The computation data model (CDM) is suggested in a form of object calculus, where DO can be represented as follows: *DO = <concept, individual, state>,* where under a *concept* a collection of functions with one and the same definition range and one and the same value range is implied. An *individual* means an entity that an expert identifies within a problem domain by specifying its unique properties. *State* changes model problem domain individual dynamics. Compared to results known as yet, principal benefits of the CDM suggested are more adequate mapping of heterogeneous weak-structured problem domain dynamics and statics as well as event-driven (meta)data control in a global computational environment. As to architecture and interface, the CDM provides "penetrating" iterative design of open, distributed, interoperable IS based on UML and BPR methodologies and .NET web-services.

A multi-level integrated IS design and implementation scheme is suggested that provides fast component-based development of integrated open extendable Web-IS with continuous (meta)data adequacy and integrity control.

The computational model is based on two-level *conceptualization* [6], i.e., the process of establishing relations between problem domain concepts.

Problem domain individuals $h$, according to assigned types $T$, are assembled into assignment-dependent collections, thus forming variable domains $H_T(I) = \{h|h : I \to T\}$, which model dynamics and statics.

When fixing CDM individuals, uniqueness of DO $d$ individualization within problem domain $D$ by formula $\Phi$ should be maintained: $||Ix\ \Phi(x)||i = d \Leftrightarrow \{d\} = \{d \in D|\ ||\Phi(d)||i=1\}$.

The DO model compression principle $C = Iy : [D]\ \forall x : D(y(x) \spadesuit \Phi) = \{x:D|\Phi\}$ is applicable to both concepts, individuals and states taken separately and to the DO on the whole.

Metadata computational model extends Codd's relational model by the compression principle: $x^{j+1} \equiv Iz^{j+1}:[...[D]...]\ \forall x^j:[...[D]...](z^{j+1}(x^j) \leftrightarrow \Phi)$, where $z^{j+1}$, $x^{j+1}$ are metalevel predicate symbols in relation to level $j$; $x^j$ is an individual of level $j$; $\Phi^j$ is a DO definition language construct of level $j$.

Hierarchical organization, scalability, metadata encapsulation, extendibility, adequacy, neutrality and semantic adequacy of the formalization provide problem-oriented IS development with (M)DO adequacy maintenance throughout the entire lifecycle.

Automated translation procedure is also developed that transforms (M)DO of the above CDM into target (M)DB schemes and (meta)data management abstract machine codes; the formal language specification presented provides data completeness, consistency and integrity.

A multi-parameter functional has been introduced: $F = F((v), (e), ...) (s) (p)$, where assignment values represent: $s$ – IS user personal preferences; $p$ – IS user registration status; $v$ – IS client interface parameters; $e$ – IS data access device parameters. Semantics object model and formal generalized procedure of (meta)data instantiation have been built, depending on the above assignments and based on functional $F$ evaluation function $||\circ||$ [8].

**2.2 Content management model**

An abstract machine for content management (AMCM) [10] is suggested as a formal model of content management IS for information resources, which is an improved version of categorical abstract machine (CAM) [2]. At any given moment AMCM is determined by its *state*. AMCM work *cycle* can be formalized by explicit enumeration of possible state changes, which define the procedure of AMCM state *dynamics* modeling.

From the formal model viewpoint, when portal page templates are mapped into the pages, variable *binding* actually occurs, which is evaluation of variables that characterize template elements with their values, or portal page elements.

AMCM semantics can be described on the basis of D.Scott semantic domain theory [5]. Therewith, atomic template types are chosen out of standard domains, while more complex template types are built using domain constructors.

Let the AMCM language contain expression set $E$ (including constant set, identifier set $I$, assignment operation (content "write operation" to template "slot") etc.), and command set $C$ (comparison, command sequence etc.).

AMCM syntax is completely defined by the following syntax domain description:
*Ide ={I | I – identifier};     Com ={C | C – command};   Exp ={E | E – expression}*.
Let us collect all possible language identifiers into *Ide* domain, commands – into *Com* domain, and expressions – into *Exp* domain.
State-based computational model of AMCM environment is as follows:
*State = Memory × Input × Output;   Memory = Ide → [Value + {unbound}];*
*Input = Value\*;      Output = Value\*;     Value = Type1 + Type2 + … .*
AMCM state is defined by "memory" state containing input values (content) and output values (web-pages) of the abstract machine. Therewith, under *memory* a mapping from identifier domain into value domain is implied, which is similar to lambda calculus variable binding. For correct exception handling, *unbound* element should be added to the domains. *Value* domain is formed by disjunctive sum of domains, which contain content types of AMCM language.
Semantic statements describe *denotates* (i.e. correct construct values) of AMCM (M)DO manipulation language.
Semantic statement for an expression reads: **E**: *Exp → [ State → [[Value × State] + {error}]]*.
Expression evaluation in AMCM environment results in such a state change that the variable is bound to its value, or (in case the binding is impossible due to variable and value type incompatibility) an error is generated.
Semantic statement for a command reads: **C** : *Com→[State→[State+{error}]]*.
Constant denotates are their respective values (in a form of ordered pair of <*variable, value*>) while program state remains unchanged.
Identifier denotates are identifiers bound with their values in case binding is possible (in a form of ordered tuples of <*variable_in_memory, identifier, state*>), while the state remains unchanged, and in case the binding is impossible an error message is generated:
**E** *[I] s = (m, I = unbound) error, → (m, I, s)*.
Semantic function for assignment command of AMCM language has the following type:
**C**: *Com → State → [State + {error}]*.
Thus, content management IS template binding with the content may result in AMCM state change and in a number of limited, predefined cases – in error generation.
Semantic statement for an AMCM command, which assigns content to template element, results in state change with substitution of content value *v* by identifier *I* in the memory:
**C** *[I=E] = **E** [E] \* λv (m , i, o) . (m [v/I], i, o)*.

## 3. IMPLEMENTATION FEATURES

The methodology has been approved by development *UniQue* HR management IS, Internet and Intranet portals in ITERA International Group of Companies.
In terms of system architecture, the IS provides assignments (depending on front-end position in data access hierarchy) with a certain level of (meta)data entry, modification, analysis and report generation. Problem-oriented form designer, report writer, online documentation and administration tools are used as interactive interface toolkit. (M)DB supports integrated storage of data (for online access) and metadata (DO dimensions, integrity constraints, data formats, etc.
During the resource management IS design, problem domain DM specifications (in terms of semantic networks) have been transformed by *ConceptModeller* to UML diagrams, then by Oracle Developer/2000 integrated CASE tool – to ERD (or by AMCM and Oracle Portal toolkit – into AM code) and, finally, into target IS and (M)DB schemes.
IS implementation process included prototyping and full-scale Oracle-based implementation. Web pages, automatically generated by information resources (meta)data management IS, have been published at ITERA Group Intranet portal and official Internet site (www.iteragroup.com).

To provide the required industrial scalability and fault tolerance, integrated Oracle IS development toolkit (Portal, Developer/2000) has been chosen to support UML and BPR technologies.

All of the components are designed, implemented and customized according to technical specifications outlined by the author and tested for several years in a heterogeneous enterprise environment. Advanced personalization and access level differentiation allows to substantially reduce risks of (meta)data damage or loss [9].

## 4. RESULTS AND CONCLUSION

A novel IS development methodology in terms of multi-level iterative procedure has been developed for information resource management, which provides adequate, consistent and integrate (meta)data manipulation during the entire lifecycle [8].

A set of models have been constructed including problem domain conceptual model and a model for development tools and computational environment in terms of state-based abstract machines. A SDK has been implemented including *ConceptModeller* visual problem oriented CASE-tool and IS for content (i.e., (M)DO) management.

To solve the task of information resource management, the full-scale enterprise IS has been customized for HR and information resource management (by Web-portals) and implemented in a corporation employing nearly 10,000 people. The obtained results proved shortening terms (1,5 times average) and reducing costs of implementation (40% average) compared to commercial software available, and high mobility, expandability, scalability and ergonomics of the IS. ITERA experts state that the implementation has resulted in annual cost reduction of hundreds of thousands of USD, while data management efficiency has increased substantially. The author is going to continue his studies of the formal models and related SDK that support enterprise content management IS.

**REFERENCES**
1. Barendregt H.P. The lambda calculus (revised edition), Studies in Logic, 103, North Holland, Amsterdam, 1984.
2. Cousineau G., Curien P.-L., Mauny M. The categorical abstract machine. Science of Computer Programming 8(2): 173-202, 1987.
3. Curry H.B., Feys R. Combinatory logic, vol.I, North Holland, Amsterdam, 1958.
4. Roussopulos N.D. A semantic network model of data bases, Toronto Univ., 1976.
5. Scott D.S. Domains for denotational semantics. ICALP 1982, 577-613.
6. Wolfengagen V.E. Event-driven objects. CSIT'99, Moscow, Russia, 1999, Vol.1.,p.p.88-96.
7. S.V.Zykov. Human Resources Information Systems Improvement: Involving Financial Systems and Other Sources Data. In: ADBIS'98, Springer, 1998, p.p. 351-356.
8. S.V.Zykov. The Integrated Approach to ERP: Embracing the Web. In: CSIT-2002, Sept., 2002. Patras, Greece.
9. S.V.Zykov. Integrating Enterprise Software Applications with Web Portal Technology. In: CSIT2003, Sept., 2003. Ufa, Russia, USATU Publishers, 2003, p.p.60-65.
10. S.V.Zykov. Abstract Machine as a Model of Content Management. CSIT'2004, Budapest, Hungary, 2004, Vol.1, p.p.251-252.
11. S.V.Zykov. Enterprise Portal: from Model to Implementation. CSIT'2004, Budapest, Hungary, 2004, Vol.2, p.p.188-193.